\documentstyle[prl,twocolumn,aps]{revtex}
\input epsf 
\input{psfig} 
\def\be{\begin{equation}} 
\def\ee{\end{equation}} 
\def\bea{\begin{eqnarray}} 
\def\eea{\end{eqnarray}} 
\def\bea*{\begin{eqnarray*}} 
\def\eea*{\end{eqnarray*}} 
\def\bit{\begin{itemize}} 
\def\eit{\end{itemize}} 
 
\begin{document} 
\title{The role of the energy gap in protein folding dynamics} 
\author{E.Pitard \thanks{% 
e-mail:pitard@spht.saclay.cea.fr} and H.Orland} 
\address{Service de Physique Th\'eorique \\ 
Centre d'Etudes de Saclay, CEA, 91191 Gif-sur-Yvette \\ 
France} 
\date{\today} 
\maketitle 
 
\begin{abstract} 
The dynamics of folding of proteins is studied by means of a 
phenomenological master equation. The energy distribution 
is taken as a truncated exponential for the misfolded 
states plus a native state sitting below 
the continuum. The influence of the gap on the folding dynamics is studied, 
for various models of the transition probabilities between the different 
states of the protein. We show that for certain models, the relaxation to 
the native state is accelerated by increasing the gap, whereas 
for others it is slowed down . 
 
%\vskip 1cm 
%\noindent\mbox{Submitted for publication to:} \hfill  
%\mbox{Saclay, SPhT/98-}\\ \noindent \mbox{``''}\\  
%\vskip 1cm\noindent \mbox{PACS: 75.10N, 64.70P, 71.55J} 
% 
\end{abstract} 
 
\pacs{\tt$\backslash$\string pacs\{\} 12.34,56.78} 
 
\pagestyle{plain} 
 
%\begin{multicols}{2}  
\section{Introduction} 
 
Proteins are known to fold into a unique native state that is biologically 
active \cite{ANF}. 
Despite their complexity and frustrated character, their folding rate is 
fast and the folding times vary from milliseconds to seconds. However, these 
times are long compared to the usual collapse time of a homopolymer (a few 
microseconds) \cite{PGG}. This means that the diversity of the aminoacids and 
the 
resulting frustration play a crucial role in the slowing down. The main 
picture that has emerged to describe this type of dynamics is that of a 
``funnel-like'' phase space \cite{FUNNEL}: the folding pathway of a protein 
in phase space 
is along a unique but rough funnel towards its ground state. However, so far, this 
picture is not based on microscopic models. 
 
Numerical simulations that have been dealing with lattice models of short 
disordered polymers tend to support this funnel picture. A number of these 
simulations seem to show that ``good folders'' (sequences that fold rapidly 
to their ground state, and are thus good candidates to represent real 
proteins) indeed follow a ``funnel'' in the energy landscape, whereas ``bad 
folders'' possess a large number of energetic traps along 
the dynamical path that 
slow down the folding process \cite{MELIN,BAN}. 
 
Several studies have tried to relate the foldability of model proteins with 
their energy gap. It has been argued in ref. \cite{SSK} 
that model proteins which 
have a large energy gap between the native state and the first 
conformationally different (compact) state fold rapidly. On the other hand, 
other studies  \cite{THI} have shown that the parameter 
which governs the rapidity of 
folding is the distance of the folding temperature to the collapse 
temperature. 
 
The aim of this paper is to show that the situation is not so simple, and 
that in particular, the folding rate depends very much on the dynamics used 
for the simulations. In particular, we show that for certain transition 
rates, a large gap accelerates the folding, whereas for other models, 
it may slow it down. 
 
\section{Modelization by a truncated REM} 
 
Many models have stressed the analogy between protein folding and the 
thermodynamics of heteropolymers \cite{REVUE}. 
In a mean field approach, some models of 
quenched disordered polymers are similar to a Random Energy Model (REM)
\cite{GO,SG}; the 
low lying states of the protein are identified with the low-lying states of 
the REM, which are responsible for the slow dynamics. Although this analogy 
is questionable as far as real proteins are concerned \cite{IRB,PANDE}, 
we will adopt this 
framework for the dynamical models studied in the following. 
 
For the REM as well as for the Sherrington Kirkpatrick (SK) model 
\cite{MPV85}, it has been shown that i) all low lying states have the same 
extensive energy and ii) the distribution of the corrections to extensivity 
of their free energies is exponential (bound at high energy but not at low 
energy). 
 
For proteins, the energies are bound below by that of the native state and 
above by that of a swollen coil. In addition, the energy landscape is known 
to be very rugged and the number of misfolded states (at fixed energy $E$) is 
known to grow very rapidly with $E$. It is thus natural, by analogy with 
many disordered systems, to assume an exponential distribution for the 
energies of the lowest lying misfolded states and to isolate the native 
state below the continuum of energies. 
 
In our model, we describe the phase space of the protein as consisting of $M$ 
misfolded states $E_\alpha$ with $\alpha=1,\ldots,M$ and one native state $% 
E_0$. The distribution of energies of the misfolded states is continuous, 
given by:  
\[ 
p(E)=\beta_c e^{\beta_c (E-E_c)}  
\] 
where $\beta_c$ is a parameter (related to the glass transition temperature 
of the REM) and $E_c$ is the energy of the highest state. 
 
More precisely, the energy levels are such that $E_{min}\leq E\leq E_c$, 
where $E_{min}$ is the energy of the first misfolded state. Therefore, 
Boltzmann weights $B_\alpha=e^{-\beta E_\alpha}$ 
vary between $B_{min}=e^{-\beta E_c}$ and $% 
B_{max}=e^{-\beta E_{min}}$. To account for the native state, we include an 
additional state with energy $E_0$ and Boltzmann weight $B_0$. 
 
The energy gap $\Delta$ is defined by $\Delta= E_{min} - E_0 >0$. 
 
The main consequence of the truncation of the energy distribution is that 
the system will never spend a very large time in one of the energy 
traps, and that $P_\alpha (t)$ will always relax towards its equilibrium 
value $P_\alpha ^{eq}$ with a finite relaxation time at large times. Slow 
dynamics features such as aging won't appear either; this is in contrast 
with the case when the distribution is not truncated, which has been studied 
extensively recently \cite{CEC}.
 
The calculations are made by taking first the limit $M\to \infty$ before 
taking the limit of large times. This is justified if the number of 
metastable conformations is large enough. 
 
The dynamics is based on the master equation: 
 
\begin{equation} 
\frac{dP_\alpha }{dt}=\sum_\beta W_{\alpha \beta }P_\beta (t)-\sum_\beta 
W_{\beta \alpha }P_\alpha (t)  \label{master} 
\end{equation} 
where $P_\alpha (t)$ is the probability of occupation of the energy level $% 
E_\alpha $ at time $t$ , and $W_{\alpha \beta }$ is the transition rate from 
the energy level $E_\beta $ to the energy level $E_\alpha $. In all the 
cases we have studied, detailed balance is satisfied:  
\begin{equation} 
\frac{W_{\alpha \beta }}{W_{\beta \alpha }}=e^{-\beta (E_\alpha -E_\beta )} 
\end{equation} 
 
When solving the master equation, the quantities that are 
usually calculated are averages over the distribution 
of disorder $p(E)$. This is 
justified in the case of a macroscopic system with short range interactions, 
but not in the case of a protein, which is too small an object for 
self-averageness to hold. 
 
This is why, in contrast with other studies, we have calculated quantities 
which are not averaged over the distribution of disorder. We considered 
three cases, depending on the choice of transition rates between energy 
levels. We find that if the transitions rates depend only on the final 
state, the relaxation is accelerated by a large gap whereas if they only 
depend on the initial state, the dynamics is slowed down; in the 
intermediate case, the situation is more complex. 
 
\section{Results} 
In this section, we present the results of the calculations, which will
be detailed elsewhere.

Studies as the ones presented below 
have already been made for a non truncated distribution of 
energies: disorder-averaged quantities show stretched exponential or 
power law
behavior at large times \cite{DO,KH}. The same models have been used
in ref. \cite{Shakh_Gutin} in the context of heteropolymer folding.

\subsection{Case where the transition rates depend only on the final state: $% 
W_{\alpha \beta }=e^{-\beta E_\alpha }$} 
 
If $E_0$ and a gap are included in the analysis, the master equation can 
easily be solved, leading to an exponential behavior,  $P_\alpha (t)=% 
\frac{e^{-\beta E_\alpha }}Z+(P_\alpha (0)-\frac{e^{-\beta E_\alpha }}Z% 
)e^{-Zt}$,
where $Z=\sum_{\alpha} B_{\alpha}$.
 Since $Z\simeq B_0 + M\bar{B_{\alpha}}=B_0+ {x \over 1-x} B_{max}$
(the bar denotes the average over the distribution of energies), 
where $B_0 = B_{max} e^{\beta \Delta}$, the 
relaxation time is $\tau ={1\over B_{max} (e^{\beta \Delta} + {x\over 1-x})}$.
 
\begin{figure}[htbp]
\begin{center}
\mbox{\epsfxsize=4cm \epsfbox{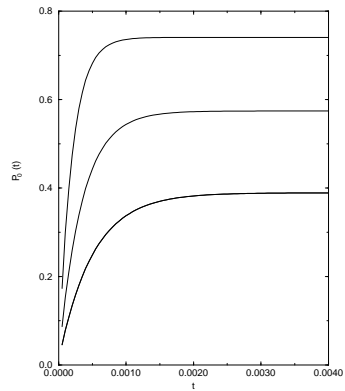}}
\caption{\small $P_0(t)$ in the case where
$W_{\alpha \beta}=e^{-\beta E_{\alpha}}$ for three sequences
with the same distribution of energies but different gaps.
From top to bottom, $\Delta=1.5$, $\Delta=1$, $\Delta=0.5$, $\beta=1.5$ and $M=100$.
The dynamics is faster for larger values of the gap.}\label{1}
\end{center}
\end{figure}

\begin{figure}[htbp]
\begin{center}
\mbox{\epsfxsize=4cm \epsfbox{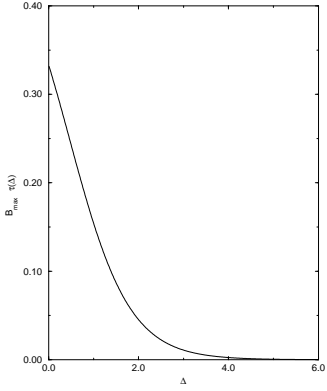}}
\caption{\small Relaxation time $\tau(\Delta)$ as a function of the gap $\Delta$
in the case where
$W_{\alpha \beta}=e^{-\beta E_{\alpha}}$, with $\beta=1.5$ and
$\beta_c=1$.}\label{2}
\end{center}
\end{figure}

Let us compare the dynamics of two protein sequences that differ only by 
their native energies $E_0$ keeping $E_{min}$ (or $B_{max}$) constant.
As seen above, the relaxation time $\tau $ decreases when 
the gap $\Delta $ increases. Indeed, in such a dynamical scheme, jumps 
towards the native state are enhanced and the lower its energy, the faster 
it is populated. The dynamics is illustrated by Fig.\ref{1} where 
the probability of occupation of the native state 
$P_0(t)$ is plotted for different values of the gap. On Fig.\ref{2}, we
show the relaxation time as a function of the gap.
\vskip0.5cm 

\subsection{Case where the transition rates only depend on the initial 
state: $W_{\alpha \beta }=\frac 1Me^{\beta E_\beta }$.} 

As was noted in ref. \cite{Shakh_Gutin}, this corresponds to the case
of a unique barrier at energy $E^*$, through which the system must
pass in order to make
a transition from state $\alpha$ to $\beta$.

Calculations can be made in both cases where the initial conditions are 
uniform (all states are equally populated) or delta-like (the initial 
probability of occupation of a specific state is one). They lead to the same 
conclusions concerning the relaxation times, and we give here the results for
the case of uniform initial conditions. At short times, the probability of 
occupation of states follows a power-law:  
\[ 
P_\alpha (t)\simeq \frac 1{x B_{max}^x}\frac 1{\Gamma (1-x)\Gamma (x)\Gamma (1+x)}t^x 
\] 
At longer times, the relaxation is exponential. For a non-native state $% 
\alpha \neq 0$, 
 
\[ 
P_\alpha (t)\simeq \frac{1-x}{x}\frac 1{B_{max}^2 e^{\beta \Delta}}
\left( \frac 1{ab}+% 
\frac 1{a(a-b)}e^{-at}-\frac 1{b(a-b)}e^{-bt}\right)  
\] 
and for the native state, 
 
\[ 
P_0(t)\simeq {1 \over 1+{x \over 1-x} e^{- \beta \Delta}} (1-e^{-bt})  
\] 
These expressions involve two time constants: 
$\tau _\alpha =\frac 1a =e^{-\beta E_\alpha }$, which is the relaxation time of the energy level 
$\alpha $ in the absence of a gap, and 
$\tau _0=\frac 1b= {B_{max} \over {1-x \over x} + e^{-\beta \Delta}} $ . 
This last relaxation time is the 
one that governs the long time dynamics of the native state. 
 
\begin{figure}[htbp]
\begin{center}
\mbox{\epsfxsize=4cm \epsfbox{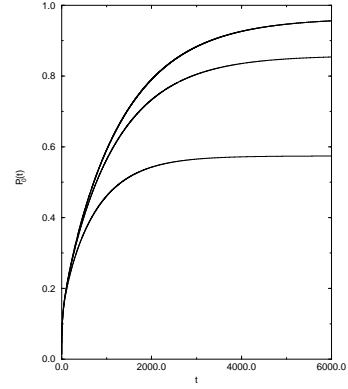}}
\caption{\small $P_0(t)$ in the case where
$W_{\alpha \beta}=e^{\beta E_{\beta}}$ for three sequences with the same
distribution of energies but different gaps.
From top to bottom, $\Delta=3$, $\Delta=2$, $\Delta=1$. $\beta=1.5$ and $M=100$.
The dynamics slows down for large values of the gap.}\label{3}
\end{center}
\end{figure}

\begin{figure}[htbp]
\begin{center}
\mbox{\epsfxsize=4cm \epsfbox{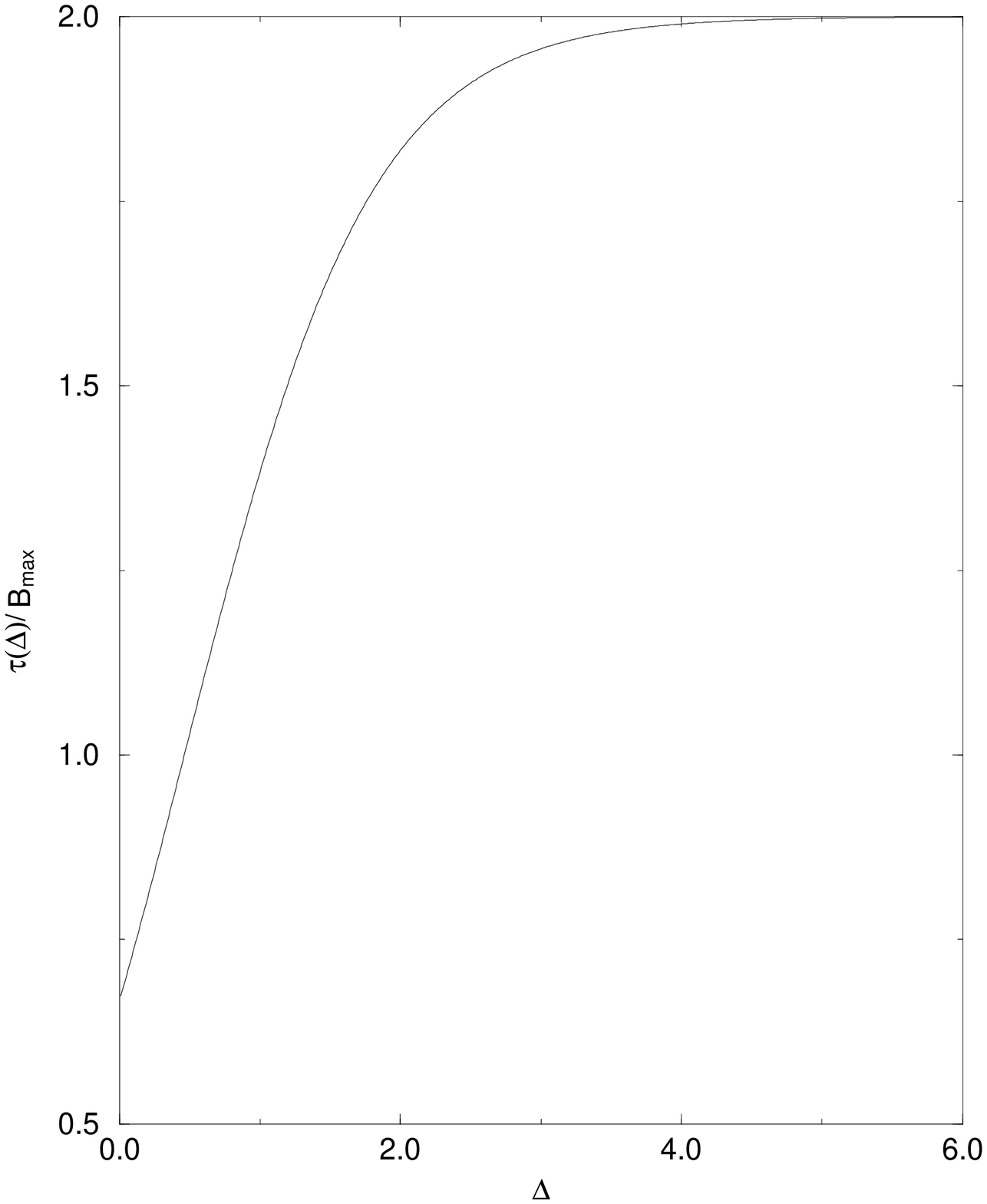}}
\caption{\small Relaxation time $\tau(\Delta)$ as a function of the gap $\Delta$
in the case where
$W_{\alpha \beta}=e^{\beta E_{\beta}}$, with $\beta=1.5$ and $\beta_c=1$.}
\label{4}
\end{center}
\end{figure}

Contrarily to the previous case, keeping $E_{min}$ (or $B_{max}$) fixed, $% 
\tau _0$ now increases as $\Delta $ increases. In this scheme, the 
energy landscape can be viewed 
as a collection of energy traps; as the energy $% 
E_0$ decreases, it becomes more difficult to escape from this state; low 
energy states are populated quite rapidly, and the system remains trapped in 
the native state; as a result, the relaxation to the Boltzmann distribution 
is slowed down. This effect is more pronounced as the energy gap gets bigger 
(see Fig.\ref{3} and \ref{4}). 
 
\vskip0.5cm 

\subsection{Intermediate case: $W_{\alpha \beta }=e^{-\beta ((1-\lambda 
)E_\alpha -\lambda E_\beta )}$} 
 
In this case, the transition rates depend both on the initial and final 
state through the above formula, where $\lambda $ is a parameter between 0 
and 1. Relaxation to the native state is described by a new relaxation time $% 
\tau _0$: 
 
\[ 
\tau _0={B_{max}^\lambda e^{\lambda \beta \Delta} \over Z} 
\frac 1{1+\frac{1-x}{x} e^{\beta \Delta}}  
\] 
where $Z=\sum_\alpha B_\alpha ^{1-\lambda }$. 
 
The dependence of $\tau _0$ on the gap is now slightly more complicated. 
One can 
distinguish two regimes: 
 
\begin{enumerate} 
\item  if $\lambda \ll 1-x $ and $\lambda \leq \frac{1}{2}$, $\tau _0$ decreases as $% 
\Delta $ increases (see Fig. \ref{5} and \ref{6}), and the relaxation is accelerated. 
On the other end, if $\lambda > \frac{1}{2}$, $\tau _0$ increases 
with $\Delta$   for $\Delta <\Delta _0$ and  decreases with 
$\Delta $ for $\Delta >\Delta _0$, where $\beta \Delta _0=
\log(\frac{2\lambda-1}{2(1-\lambda)}\frac{x}{1-x})$.

\begin{figure}[htbp]
\begin{center}
\mbox{\epsfxsize=4cm \epsfbox{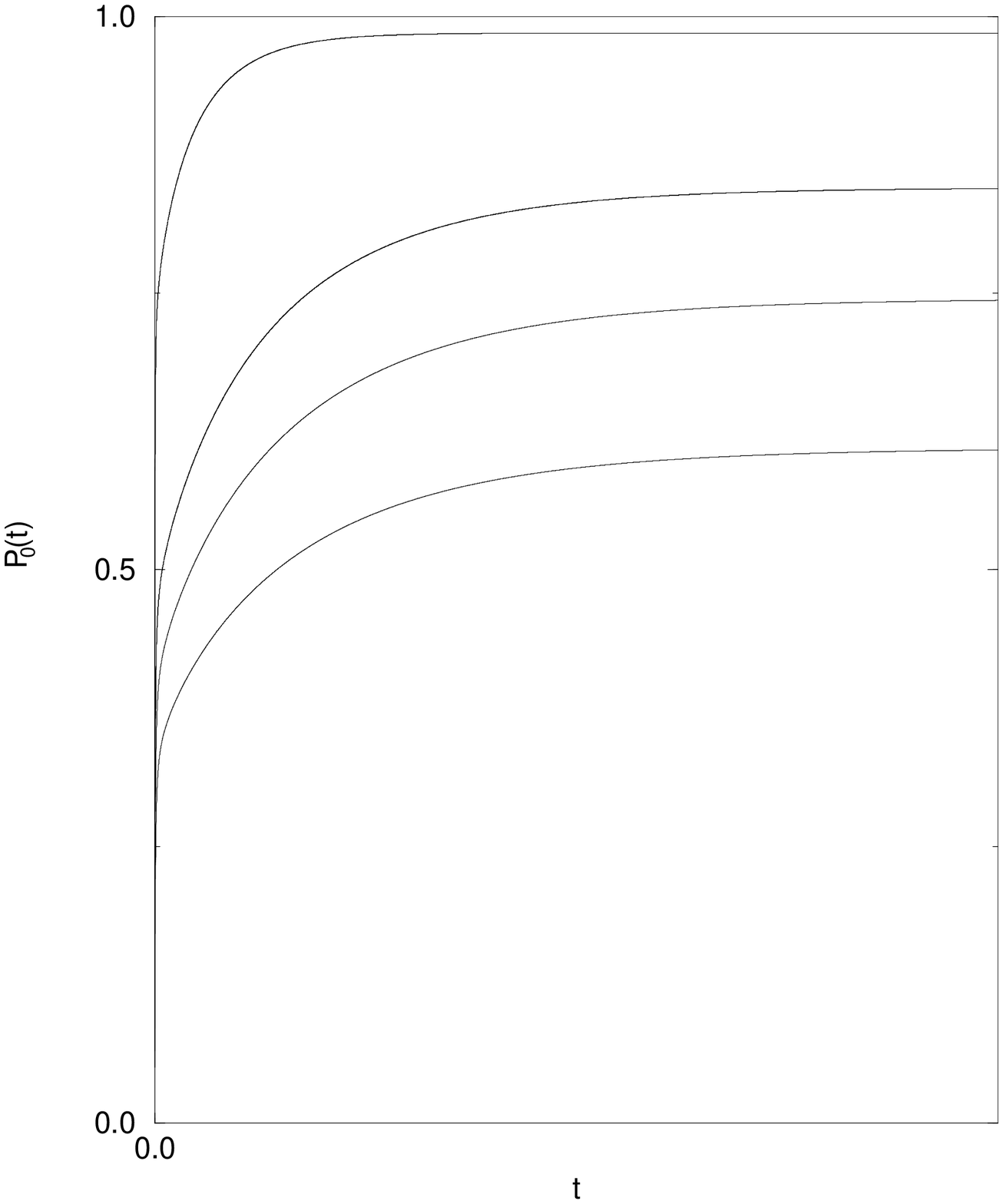}}
\caption{\small $P_0(t)$ in the case where
$W_{\alpha \beta}=e^{-\beta ((1-\lambda)E_{\alpha}- \lambda E_{\beta})}$ 
for four sequences with the same distribution of energies but different gaps. From top to bottom, $\Delta=2$, $\Delta=1$, $\Delta=0.75$, $\Delta=0.5$. We have chosen
$\lambda=0.5$, $\beta=2.5$ and $M=100$.
For this set of parameters, the dynamics is faster as the gap increases.}\label{5}
\end{center}
\end{figure}

\begin{figure}[htbp]
\begin{center}
\mbox{\epsfxsize=4cm \epsfbox{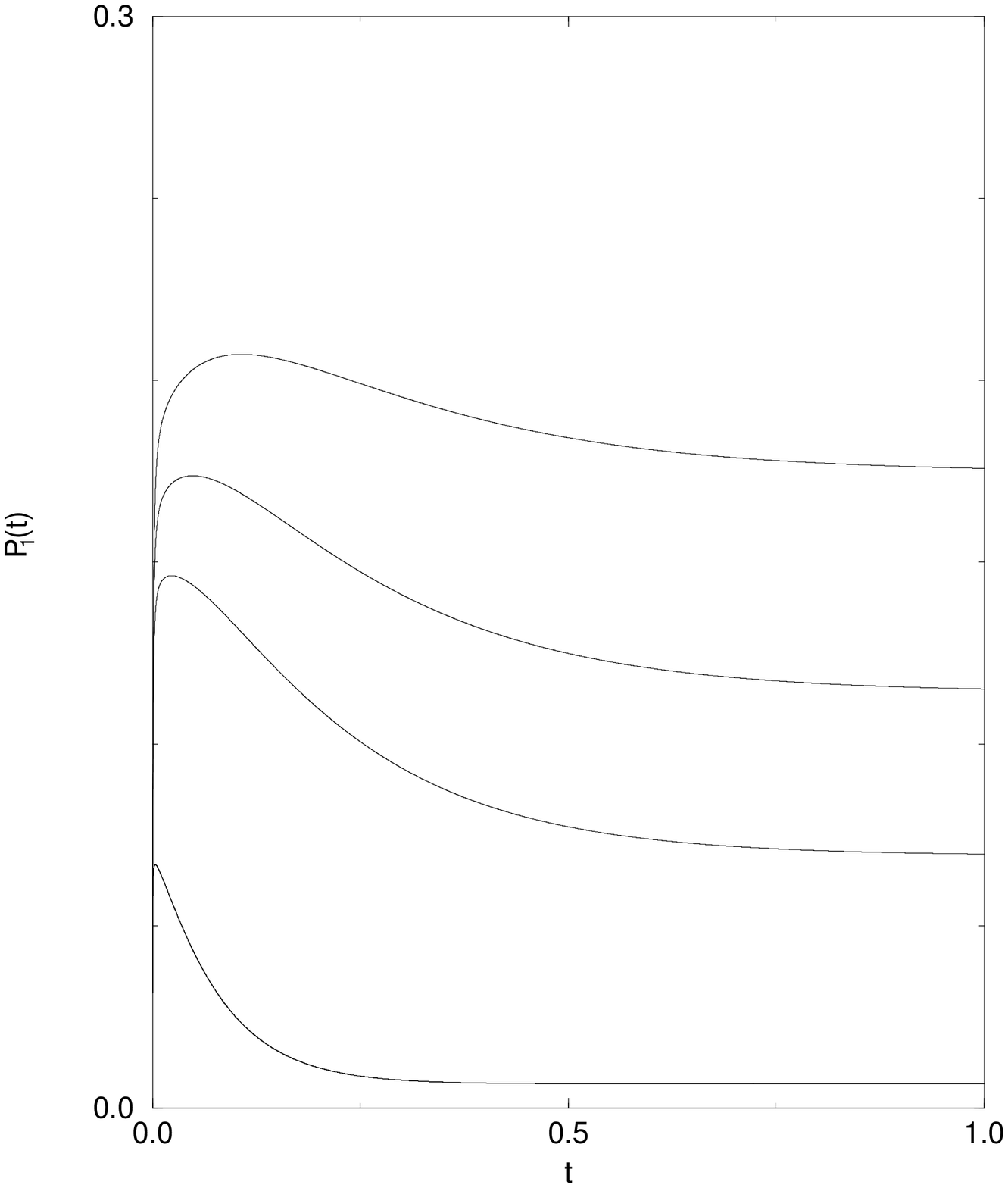}}
\caption{\small $P_1(t)$ in the case where
$W_{\alpha \beta}=e^{-\beta ((1-\lambda)E_{\alpha}- \lambda E_{\beta})}$ for four sequences with the same distribution of energies but different gaps. From top to bottom, $\Delta=0.5$, $\Delta=0.75$, $\Delta=1$, $\Delta=2$. We have chosen
$\lambda=0.5$, $\beta=2.5$ and $M=100$.}\label{6}
\end{center}
\end{figure}
\
\item  if $\lambda \gg 1-x$, $\tau _0$ is an increasing 
function of $\Delta $ for $\Delta <\Delta _0$ and a decreasing function of $% 
\Delta $ for $\Delta >\Delta _0$, where $\beta \Delta _0=\log 
(\frac {x \lambda}{\ (1-x) (1-\lambda )})$. 
%Relaxation slows down as the gap increases. 
\end{enumerate}

\section{Discussion} 
 
It seems difficult to compare our results with experiments in real proteins, 
since no systematic study of the gap of given sequences has been carried out 
experimentally. However, there has been a number of lattice simulations 
that lead to a variety of results and interpretations. Klimov and Thirumalai  
\cite{THI} claim that there seems to be no direct correlation between the 
gap and the folding dynamics. 
 
On the other hand, Sali, Shakhnovich and Karplus \cite{SSK} have shown that 
the ``best'' folders are those with the largest energy gap. In all cases, 
the simulations are performed on very short chains (27 monomers at best) and 
the interactions between monomers are random with a Gaussian distribution. 
The energy of a configuration is given by  
\[ 
E=\sum_{i<j}\Delta (r_i,r_j)B_{ij} 
\] 
where $\Delta (r_i,r_j)=1$ if monomers $i$ and $j$ are neighbors on the 
lattice and  
\[ 
P(B_{ij})=\frac 1{(2\pi B^2)^{\frac 12}}\exp \left( -\frac{(B_{ij}-B_0)^2}{% 
2B^2}\right)  
\] 
with a negative parameter $B_0$ in order to mimic the hydrophobic character 
of the solvent. These previous studies are applicable only to very short 
proteins. For longer chains, there are no results indicating how a protein 
will find its folding path through its complicated energy landscape. Some 
attempts have been made to understand analytically the dynamics of random 
heteropolymers \cite{SH-dyn} \cite{TH-dyn}; 
and at present, 
the debate regarding the interpretation of the simulations \cite{THI} \cite 
{SSK} \cite{MELIN} \cite{REPLY}  is still open. 
 
%; one can question the validity of using a Gaussian distribution for the interactions.  
%A more realistic energy landscape would be obtained by using the Miyazawa-Jernigan  
%matrix \QCITE{cite}{}{MI-JE} for contact energies between aminoacids;  
%it has already been pointed out that the interaction potential between monomers of a  
%protein is not random \QCITE{cite}{}{IRB}. 
 
Our phenomenological approach based on a REM energy landscape, although far 
from realistic, might capture the long time relaxation laws of random 
chains. The finiteness of objects such as proteins is taken into account by 
truncating the energy spectrum. Moreover, we show that the dynamics depends 
on the whole energy landscape, rather that only on one parameter, such as the 
gap. 
 
According to the type of transition rates used in the dynamics, one can 
obtain different behaviors for the relaxation as a function of the gap. It 
seems difficult to decide which one is more realistic.
It would 
be interesting to have systematic results on the dynamics of synthesized 
sequences, generated for example through mutation experiments \cite{OAS}.

%\end{multicols} 
\end{document}